\documentclass[twocolumn,prb,showpacs]{revtex4}%
\usepackage{graphicx}%
\usepackage{amsmath}%
\usepackage{amsfonts}%
\usepackage{amssymb}
\usepackage{bm}

\def\s{{\sigma}}
\def\e{{\epsilon}}
\def\k{{ {\bm k} }}

\def\q{{ {\bm q} }}
\def\Q{{ {\bm Q} }}

\def\w{{\omega}}
\def\a{{\alpha}}

\allowdisplaybreaks[4]

\begin{document}
\title{
Non-Fermi-Liquid Transport Phenomena and Superconductivity
Driven by Orbital Fluctuations in Iron Pnictides:
Analysis by Fluctuation-Exchange Approximation
}
\author{Seiichiro \textsc{Onari}$^{1}$,
and Hiroshi \textsc{Kontani}$^{2}$}
\date{\today }

\begin{abstract}
We study the five-orbital Hubbard model including the charge quadrupole
 interaction for iron pnictides.
Using the fluctuation-exchange approximation, 
orbital fluctuations 
evolve inversely proportional to the temperature,
and therefore the resistivity shows linear or convex $T$-dependence 
for a wide range of temperatures.
We also analyze the Eliashberg gap equation, and show that an
$s$-wave superconducting state without 
sign reversal ($s_{++}$-wave state) emerges
when the orbital fluctuations dominate over the spin fluctuations.
When both fluctuations are comparable, their competition gives
rise to a nodal $s$-wave state.
The present study offers us a unified explanation
for both the normal and superconducting states.
\end{abstract}
%\draft

\address{
$^1$ Department of Applied Physics, Nagoya University and JST, TRIP, 
Furo-cho, Nagoya 464-8602, Japan. 
\\
$^2$ Department of Physics, Nagoya University and JST, TRIP, 
Furo-cho, Nagoya 464-8602, Japan. 
}
 
\pacs{74.20.-z, 74.20.Fg, 74.20.Rp}

\sloppy

\maketitle

%%%%%%%%%%%%%%%%%%
%Introduction
%%%%%%%%%%%%%%%%%%
\section{Introduction}
The many-body electronic states and the pairing mechanism 
in iron pnictides have been significant open problems.
By taking account of the Coulomb interaction 
and the nesting of the Fermi surfaces (FSs) in Fig. \ref{fig1-4}(a),
a fully-gapped sign-reversing $s$-wave state ($s_\pm$-wave state) 
had been proposed 
%based on the spin fluctuation theories
 \cite{Kuroki,Mazin,Tesanovic,Hirschfeld,Chubukov}.
%However, $s_\pm$-wave superconducting (SC) state is fragile 
%against impurities \cite{Onari-impurity}, 
%inconsistently with experiments \cite{Sato-imp}.
Experimentally,
both $T_{\rm c}$ and antiferro (AF) spin correlation
increases as $x$ decreases in BaFe$_2$(As$_{1-x}$P$_x$)$_2$ \cite{Nakai}.
In contrast, $T_{\rm c}$ in LaFeAsO$_{1-x}$F$_x$ at $x=0.14$ 
increases from 23 K to 43 K by applying the pressure, whereas
AF spin correlation is almost unchanged \cite{Fujiwara}.
Thus, the relationship between $T_{\rm c}$ and strength of the
spin fluctuation seems to depend on compounds.

On the other hand, an orbital-fluctuation-mediated $s$-wave state 
without sign reversal ($s_{++}$-wave state) had been proposed 
based on the five-orbital Hubbard model including the charge quadrupole
 interaction
\cite{Onari-VC,Kontani-Onari,Saito,Saito2,Kontani-quad}.
%add
The charge quadrupole interaction is induced by the vertex correction
(VC)\cite{Onari-VC} due to the Coulomb interaction and the electron-phonon ($e$-ph) interaction due to Fe-ion Einstein oscillations.
Within the random-phase-approximation (RPA),
it was found that 
$d$-orbital fluctuation is induced by small $e$-ph interaction.
Especially, the empirical relationship between $T_{\rm c}$ and the As-Fe-As bond angle
(Lee plot) \cite{Lee} has been naturally explained.
Recently, theoretically predicated orbital fluctuations 
\cite{Kontani-Onari,Saito}
have been detected via the substantial softening of the shear modulus
 \cite{Yoshizawa}.
The softening of the shear modulus and the structure transition have been explained by the two-orbiton mechanism based on the orbital fluctuation theory\cite{Kontani-quad}.
The $s_{++}$-wave state is consistent with the robustness of $T_{\rm c}$
against randomness \cite{Onari-impurity,Sato-imp,Nakajima} 
as well as the ``resonance-like'' peak structure 
in the neutron inelastic scattering \cite{Onari-resonance}.
%The $s_{++}$-wave state is consistent with the robustness of $T_{\rm c}$ 
%against various impurities \cite{Sato-imp},
%as shown in Ref. \cite{Onari-impurity}.
%Note that the ``resonance-like'' peak structure 
%in neutron inelastic scattering \cite{christianson} 
%is reproduced theoretically even in the $s_{++}$-wave state,
%by considering the inelastic quasiparticle damping 
% \cite{Onari-resonance}.

However, spin/orbital fluctuations obtained by the RPA are
reduced by the self-energy correction.
Therefore, in order to confirm the orbital fluctuation scenario,
it is desired to analyze the many-body electronic states beyond the RPA.
For this purpose, the fluctuation-exchange (FLEX) approximation \cite{Bickers}
would be appropriate, in which the absence of spin/orbital order in 2D systems, 
known as the Mermin-Wagner theorem, is rigorously satisfied \cite{Mermin-Wagner}.

In this paper, we analyze the five-orbital Hubbard model including the
charge quadrupole interaction
for iron pnictides using the FLEX approximation.\cite{preprint}
In the normal state, large orbital fluctuations induce
highly anisotropic quasiparticle lifetime 
%(hot/cold spot structure) 
on the FSs as well as 
the $T$-linear or $T$-convex resistivity $\rho$
 \cite{Hall,Kasahara-RH,Eisaki} and the large negative thermo-electric power $S$.
The large orbital fluctuations also introduce
the $s_{++}$-wave superconducting (SC) state for a wide range of parameters,
and the competition between orbital and spin fluctuations
lead to the nodal $s$-wave state.
We propose that the orbital fluctuation is the origin of 
both the $s_{++}$-wave SC state and the
non-Fermi-liquid transport phenomena in the normal state.

\section{Formulation}
In this paper, we set the $x$ and $y$ axes parallel to the nearest Fe-Fe
bonds and the orbital $z^2$, $xz$, $yz$, $xy$ and
$x^2-y^2$ orbitals are denoted as 1, 2, 3, 4 and 5, respectively.

\begin{figure}[!htb]
\includegraphics[width=0.9\linewidth]{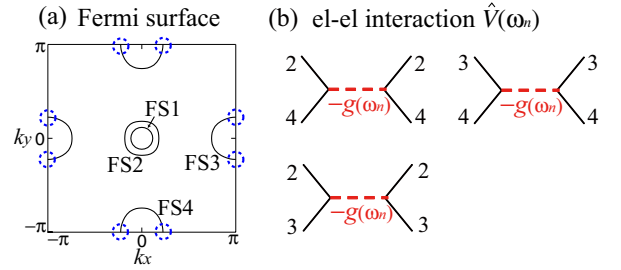}
\caption{(color online) (a) FSs in the unfolded zone.
The dotted circles represent the cold-spot given by the 
orbital fluctuation theory.
The cold-spot is composed of $xz/yz$-orbitals.
(b) Phonon-mediated el-el interaction ($\hat{V}$) for $2,3,4$ orbitals.}
\label{fig1-4}
\end{figure}

We employ the five-orbital Hubbard model\cite{Kuroki} including the
quadrupole-quadrupole [electron-electron (el-el)] interaction
induced by the VC due to the Coulomb interaction and $e$-ph interaction due to Fe-ion Einstein optical modes.
The quadrupole-quadrupole interaction is given
as\cite{Kontani-quad}
\begin{eqnarray}
\hat{V}(\omega_n)=-g(\omega_n)\sum_i\left(\hat{O}^i_{xz}\hat{O}^i_{xz}+\hat{O}^i_{yz}\hat{O}^i_{yz}+\hat{O}^i_{xy}\hat{O}^i_{xy}\right),
\end{eqnarray}
%\begin{eqnarray}
%H_{e{\rm -ph}}=\eta\sum_i\left(\hat{O}^i_{yz}u^i_x+\hat{O}^i_{xz}u^i_y+\hat{O}^i_{xy}u^i_z\right),
%\end{eqnarray}
where $\hat{O}^i_{\Gamma}$ $(\Gamma=xz,yz,xy)$ is the charge quadrupole
operator and $g(\omega_n)=g\omega_{\rm D}^2/(\omega_n^2+\omega_{\rm D}^2)$
is proportional to the phonon Green function;
$g=g(0)$ is the effective el-el interaction for $\w_n=0$, and 
$\omega_{\rm D}$ is the cutoff frequency \cite{Kontani-Onari}.
For example, we show non-zero $V_{ll',mm'}$ for $l,l',m,m'=2,3,4$
in Fig. \ref{fig1-4}(b).
Other than Fig. \ref{fig1-4}(b),
$\hat{V}$ has many non-zero off-diagonal elements
as explained in Ref. \cite{Saito,Kontani-quad},
since the Fe-ion oscillation (non-A$_{1g}$ mode)
induces various inter-orbital transitions.

In the FLEX approximation \cite{Bickers},
the $5\times5$ self-energy matrix $\hat{\Sigma}$
in the orbital representation is given by
\begin{equation}
\Sigma_{l_1l_3}(k)=\frac{T}{N}\sum_q\sum_{l_2l_4}V_{l_1l_2,l_3l_4}^\Sigma(q)G_{l_2l_4}(k-q),
\label{eqn:Sigma}
\end{equation}
where $l_i$ represents the orbital, $N$ is the number of $\bm{k}$ meshes,
and we denote $k=(\bm{k},\epsilon_n)$ with fermion Matsubara frequency
$\epsilon_n=(2n+1)\pi T$, and $q=(\bm{q},\omega_n)$ with boson Matsubara
frequency $\omega_n=2n\pi T$.
%add

%\begin{eqnarray}
%\hat{O}^i_\Gamma=\sum_{lm}o^{l,m}_{\Gamma}\hat{m}^i_{l,m},
%\end{eqnarray}
%where $\hat{m}^i_{l,m}=\sum_\sigma c^\dagger_{i,l\sigma}c_{i,m\sigma}$.
%The non-zero coefficients are given as

$\hat{G}$ is the $5\times 5$ Green function matrix in the orbital basis,
and $\hat{V}^\Sigma$ is the $25\times 25$ interaction term
for the self-energy given as \cite{Takimoto}
\begin{eqnarray}
\hat{V}^\Sigma(q)&=&\frac{3}{2}\hat{\Gamma}^s\hat{\chi}^s(q)\hat{\Gamma}^s+\frac{1}{2}\hat{\Gamma}^c\hat{\chi}^c(q)\hat{\Gamma}^c \nonumber\\
&-&\frac{1}{4}(\hat{\Gamma}^s-\hat{\Gamma}^c)\hat{\chi}^{\rm irr}(q)(\hat{\Gamma}^s-\hat{\Gamma}^c)+\frac{3}{2}\hat{\Gamma}^s+\frac{1}{2}\hat{\Gamma}^c,
\label{eff}
\end{eqnarray}
where the irreducible susceptibility is given by
\begin{equation}
\chi^{\rm irr}_{l_1l_2,l_3l_4}(q)=-\frac{T}{N}\sum_kG_{l_1l_3}(k+q)G_{l_4l_2}(k),
\end{equation}
and the spin (orbital) susceptibility is obtained as
\begin{equation}
\hat{\chi}^{s(c)}=\frac{\hat{\chi}^{\rm
 irr}}{1-\hat{\Gamma}^{s(c)}\hat{\chi}^{\rm irr}}.
\end{equation}
Here, $\hat{\Gamma}^{s}=\hat{S}$ [$\hat{\Gamma}^{c}=-\hat{C}-2\hat{V}(\w_n)$]
is the irreducible vertex for the spin [charge] channel;
%$\hat{\Gamma}^{s}=\hat{S}$ and $\hat{\Gamma}^{c}=-\hat{C}-2\hat{V}$.
${\hat S}$ and ${\hat C}$ represent the Coulomb interaction 
in the multiorbital model introduced in Refs. 
 \cite{Kuroki,Takimoto,Kontani-Onari,Saito};
Their matrix elements consist of
the intra-orbital Coulomb $U$, the inter-orbital Coulomb $U'$, Hund's
coupling $J$ and the pair hopping $J'$.
We assume that $J=J'$ and $U=U'+2J$.
Since the Fe-ion oscillation induces various inter-orbital transitions,
the substantial orbital fluctuations appear
at low frequencies. On the other hand, the charge susceptibility 
$\chi^c(\bm{q})=\sum_{lm}\chi^c_{ll,mm}(\bm{q})$ is not enhanced 
due to the cancellation \cite{Kontani-Onari,Saito}.
In the present study, we drop ladder-type diagrams by $\hat{V}(\w_n)$, 
which is justified when $\w_{\rm D}\ll E_{\rm F}$ \cite{Kontani-Onari,Saito}.
For the same reason, $\hat{V}(\w_n)$ is absent in ${\hat \Gamma}^s$.

%In the calculation, we include the all $e$-ph vertices made by the five orbitals in $\hat{V}$.

In the FLEX approximation,
we obtain $\hat{G}$ and $\hat{\Sigma}$ self-consistently
using the Dyson equation $\hat{G}^{-1}=(\hat{G}^0)^{-1}-\hat{\Sigma}$.
%where $\hat{G}^0$ is the non-interaction Green function.
In multiband systems, the FSs are modified from the original FSs
due to the self-energy correction.
To escape from this difficulty, we subtract the constant term
$[\hat{\Sigma}(\bm{k},+i0)+\hat{\Sigma}(\bm{k},-i0)]/2$
from the original self-energy, corresponding to the
elimination of double-counting terms between LDA and FLEX
\cite{Ikeda}.
%\cite{Yamada,Ikeda}.
Hereafter, 
we fix $J/U=1/6$, $\omega_{\rm D}=0.02$eV, and the electron filling
$n=6.1$ except for Fig. \ref{fig3}.
Because of the smallness of the FSs in Fig. \ref{fig1-4},
fine $\bm{k}$ meshes are required for a quantitative study.
We take $N=128\times128$ $\bm{k}$ meshes 
which is four times that used in Ref. \cite{Ikeda},
and 1024 Matsubara frequencies.
Then, we obtain reliable numerical results for $T\ge0.01$eV.

%%%%%%%%%%%%%%%%%%%%%%%%%%%%%%%%%
%%%%%%%%%%%%%%%%%%%%%%%%%%%%%%%%%
\begin{figure}[!htb]
\includegraphics[width=\linewidth]{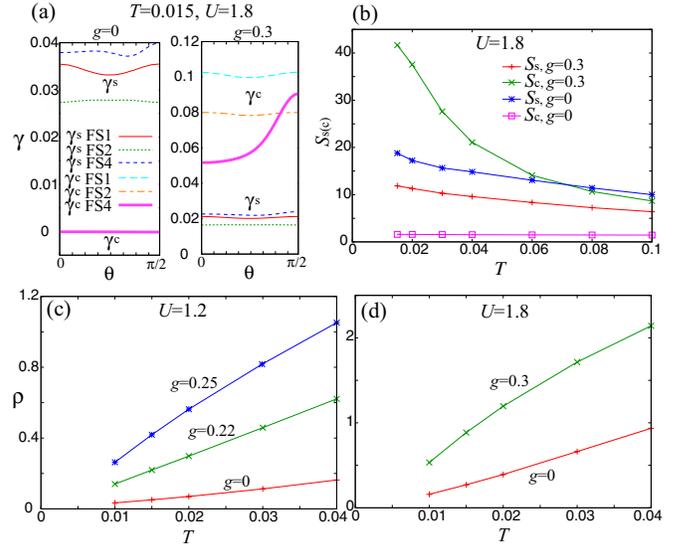}
\caption{(color online) 
(a) $\k$-dependence of $\gamma^{s(c)}$ induced by the spin (orbital)
 fluctuations on each FS.
Note that $\gamma^s$ decreases with $g$
due to the suppression of $\hat{\chi}^s$ by $\gamma^c$.
%, where $\theta$
% denotes the azimuthal angle measured from $\Gamma$(M) point for
% FS1,2(4), and $\theta=0$ corresponds to the $k_x$ direction.
%which is mainly composed of 2,3 orbitals. (4 orbital).
(b) $T$-dependence of $S_{s(c)}=(1-\a_{s(c)})^{-1}$. 
(c),(d) $T$-dependence of $\rho$.
$\rho=1$ corresponds to $(\hbar a_c)/e^2\sim 300\mu$cm$\Omega$
when the interlayer distance is $a_c=0.6$nm.
}
\label{fig1}
\end{figure}
%%%%%%%%%%%%%%%%%%%%%%%%%%%%%%%%%

\section{Result}
\subsection{Normal state}
We begin with the electronic property in the normal state.
Hereafter, the unit of energy is eV.
First, we discuss the quasiparticle damping rate
$\gamma_{\bm k}$ on each FS, which is given by the imaginary part 
of the self-energy in the band-diagonal representation.
%$\gamma_{\bm k}=-{\rm Im}\Sigma(\bm{k},0)$
%which is the quasi-particle damping without mass renormalization on each FS for
%$T=0.015$, $U=1.8$. $\gamma^{s(c)}_{\bm{k}}$ induced by the spin (orbital)
In Fig. \ref{fig1}(a), 
$\gamma^{s(c)}_{\bm{k}}$ represents the damping due to 
spin (orbital) fluctuations for $T=0.015$ and $U=1.8$,
which is given by substituting
$\hat{V}^\Sigma=\frac{3}{2}\hat{\Gamma}^s\hat{\chi}^s\hat{\Gamma}^s+\frac{3}{2}\hat{\Gamma}^s$
$(\frac{1}{2}\hat{\Gamma}^c\hat{\chi}^c\hat{\Gamma}^c+\frac{1}{2}\hat{\Gamma}^c)$ 
in Eq. (\ref{eqn:Sigma}).
The horizontal axis is the azimuth angle for the $\bm{k}$ point 
with the origin at the $\mathrm{\Gamma}$($M$) point for FS1,2 (FS4).
%$\theta = 0$ corresponds to the $k_x$ direction.
The relationship $\gamma_\k\approx\gamma^{s}_\k+\gamma^{c}_\k$ is satisfied 
since the third term in Eq. (\ref{eff}) is very small.
%add
We will see below that the value $U=1.8$ can reproduce moderate
AF spin fluctuations observed in $e$-doped compounds, and
it is consistent with $U\sim2$ reported by 
x-ray absorption spectroscopy (XAS) \cite{x-ray}.

In Fig. \ref{fig1}(a),
the relation $\gamma^s\gg\gamma^c$ holds for $g=0$,
%since \the orbital fluctuation is very small.
and the momentum dependence of $\gamma_\k^s$ on each FS is small
although the AF spin correlation is well developed.
The value of $\gamma^c$ increases with $g$,
and $\gamma^c\sim\gamma^s$ at $g=0.26$.
In Fig. \ref{fig1}(a), $\gamma^c\gg\gamma^s$ for $g=0.3$;
the corresponding dimensionless coupling is just $\lambda\equiv gN(0)\sim0.2$.
%(We verified that $\gamma^s\sim\gamma^c$ at $g=0.26$.)
%Since $\gamma_\k^c$ is mainly induced by the 
%orbital fluctuations between $2(3)$- and $4$-orbitals
% $\chi_{24,24}(\Q)$
Then, $\gamma_\k^c$ on FS4 (e-pocket) is anisotropic
due to the orbital dependence of $\chi_{ll',mm'}$, and
%shows large $\theta$-dependence; it 
it takes the minimum value at $\theta\sim0$,
where the FS is composed of $2,3$-orbitals \cite{Kuroki}.
This ``cold-spot'' is important for the transport phenomena.
Since the cold spot is on the e-pocket,
the Hall coefficient $R_{\rm H}$ will be negative,
consistent with experiments
\cite{Sato-RH,Hall,Kasahara-RH}.
In the case of high-$T_{\rm c}$ cuprates, 
various non-Fermi-liquid transport phenomena 
(e.g., violation of Kohler's rule) originate from the 
cold/hot spot structure as well as the backflow 
(=current vertex correction) due to the spin fluctuations
\cite{ROP}.
By analogy, the appearance of the cold spot in Fig. \ref{fig1}(a)
indicates that the orbital fluctuations are the origin of 
striking non-Fermi-liquid transport phenomena in iron pnictides
\cite{Sato-RH,Hall,Kasahara-RH}.

In Fig. \ref{fig1}(b), we show how the orbital and spin fluctuations 
develop as $T$ decreases:
In the FLEX, the spin (orbital) susceptibility is enhanced by the 
spin (orbital) Stoner enhancement factor $S_{s(c)}=(1-\alpha_{s(c)})^{-1}$, 
where $\a_{s(c)}$ is the maximum of the largest eigenvalue of 
$\hat{\Gamma}^{s(c)}\hat{\chi}^{\rm irr}(\bm{q},0)$ with respect to $\bm{q}$.
Then, $\alpha_{s,c}=1$ corresponds to the spin/orbital order, although
it is prohibited in 2D systems by the Mermin-Wagner theorem
 \cite{Mermin-Wagner}.
In the case of $U=1.8$ and $g=0$,
large $S_{s}\ (\gtrsim10)$ is produced at $\q\approx\Q\equiv(\pi,0)$
(i.e., $\chi^s(\Q,0)\propto S_{s}$).
$S_{s}$ gradually increases as $T$ drops, which is a typical critical 
behavior near the AF magnetic quantum-critical-point (QCP)
\cite{Moriya}.
When $g>0$, $\chi^c(\bm{q},0)$ is enhanced 
at $\q=\bm{0}$ and $\q=\Q$ almost equivalently \cite{Saito}.
At $g=0.3$, large $S_{c}\ (\gg10)$ is produced at 
$\q\approx\Q$ or ${\bm 0}$,
and it increases approximately proportional to $T^{-1}$.
Thus, it is confirmed that both ferro- and AF-orbital fluctuations 
show critical evolutions near the orbital QCP.

Next, we discuss the resistivity $\rho$ due to 
the orbital and spin fluctuations.
By neglecting the backflow, the conductivity is obtained by
\begin{eqnarray}
%\sigma_{xx}=\frac1N \sum_{\bm{k},\alpha}\int
% \frac{d\omega}{\pi}\left(-\frac{\partial f(\omega)}{\partial \omega}\right)
% \frac{{v^x_{\a,\k}}^2}{(\w-\e^0_{\a,\k})^2+\gamma_{\a,\k}^2(\w)},
% \nonumber
\sigma_{xx}=\frac{e^2}{N} \sum_{\bm{k},\alpha}\int_{-\infty}^{\infty}
 \frac{d\omega}{\pi}\left(-\frac{\partial f(\omega)}{\partial \omega}\right)
 \left|v^x_{\a,\k} G_{\k,\a}(\w+i0)\right|^2 ,
\end{eqnarray}
where $e(<0)$ is the charge of an electron, $\a$ is the band index,
$f(\omega)$ is the Fermi distribution function, 
$v^x_{\a,\k}$ is the velocity of band $\a$,
and $G_{\k,\a}(\w+i0)$ is the retarded Green function 
for band $\a$ in the FLEX approximation.
%$\gamma_{\a,\k}$ is the quasiparticle damping 
%given by the FLEX approximation,
%$v^x_{\a,\k}={\partial \epsilon^0_{\a,\k}}/{\partial k_x}$,
%and $\epsilon^0_{\a,\k}$ is the band dispersion
%measured from the Fermi level.
Figure \ref{fig1}(c) and (d) show the obtained resistivity $\rho=1/\s_{xx}$
for $U=1.2$ and 1.8:
In case of $U=1.2$, 
$\rho$ shows a conventional sublinear (concave) $T$-dependence at $g=0$.
%Since $\rho\propto \gamma^s+\gamma^c$ at the cold spot, 
%$\rho$ increases with $g$, and
%almost $T$-linear resistivity is realized at $g=0.2$.
$\rho$ increases with $g$ due to the orbital fluctuations, and
almost $T$-linear resistivity is realized at $g=0.22$.
At $g=0.25$, $\rho$ shows a superlinear (convex) $T$-dependence.
In case of $U=1.8$,
$\rho$ is linear-in-$T$ at $g=0$, while it 
shows a clear superlinear $T$-dependence at $g=0.3$.
% stress that
%we should keep in mind that
Thus, we stress that
non-Fermi-liquid resistivity is not a direct evidence for 
the spin fluctuations.
In $Ln$FeAsO compounds,
$T_{\rm c}$ increases as the As$_4$ tetrahedron is close to a regular one,
%as the radius of lanthanide ion $Ln^{3+}$ decreases,
and the $T$-dependence of $\rho$ changes from concave
% (low $T_{\rm c}$) 
to convex
% (high $T_{\rm c}$) 
\cite{Eisaki}.
Since $g$ is maximum when the As$_4$ tetrahedron is regular \cite{Saito},
this experimental correlation between $T_{\rm c}$ and $\rho(T)$
is understood in terms of the orbital fluctuation scenario. We note
that non-Fermi-liquid-like frequency dependence of Im$\Sigma(\w)$
was recently discussed in Ref. \cite{Lee-NFL}.
\begin{figure}[!htb]
\includegraphics[width=0.9\linewidth]{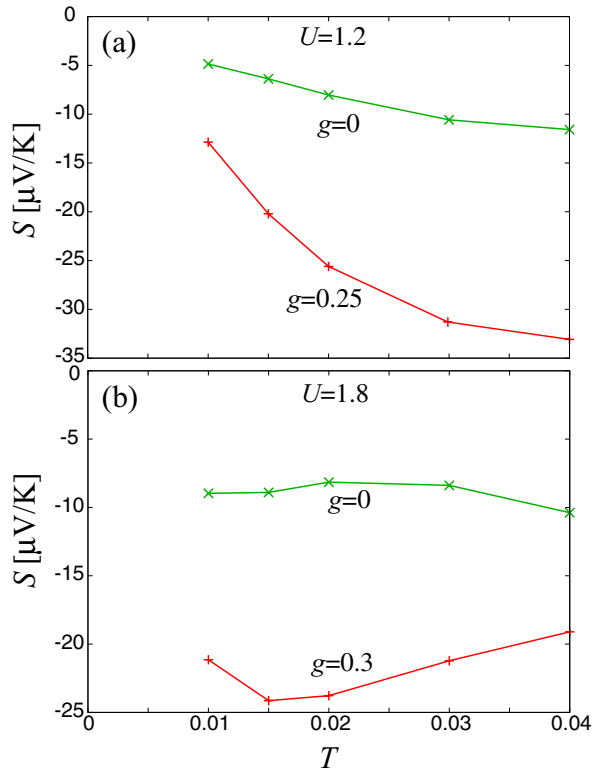}
\caption{(Color online) 
$T$-dependence of $S$ for (a) $U=1.2$ and (b) $U=1.8$
}
\label{fig5}
\end{figure}

Here, we study the non-Fermi-liquid-like behavior of the thermo-electric power $S$ induced by the orbital and
spin fluctuations. According to experimental
results\cite{Mcguire,Tropeano,Singh,Awana,Mun}, $S$ in $e$-doped
systems is negative below the room temperature, and $|S|$ develops inversely
proportional to the temperature till $T^*\sim 100$K.
In optimum doped Ba(Fe$_{1-x}$Co$_x$)$_2$As$_2$ peak value of
$|S|\sim50\mu$V/K and $T^*\sim130$K are observed.\cite{Mun} 
Similar behaviors are also observed in high-$T_c$ cuprates with the strong
AF spin fluctuation.
Thus, such remarkable non-Fermi-liquid behaviors in iron pnictides are
expected to be realized by the orbital and spin fluctuations.

Since the effect of backflow is small
for $S$,\cite{Kontani-S} we calculate $S$ by neglecting the backflow as follows
\begin{eqnarray}
S=\frac{e}{\sigma_{xx}TN} \sum_{\bm{k},\alpha}\int_{-\infty}^{\infty}
 \frac{d\omega}{\pi}\left(-\frac{\partial f(\omega)}{\partial \omega}\right)
 \omega\left|v^x_{\a,\k} G_{\k,\a}(\w+i0)\right|^2.
\end{eqnarray}
The obtained $T$-dependence of $S$ is shown in Fig. \ref{fig5}.
In the case of $U=1.2$, Fermi-liquid-like behavior $(S\propto T)$ is
obtained for $g=0$, where both the spin and orbital fluctuations are weak.
For $g=0.25$, where the orbital fluctuation is strong, $|S|$ becomes
larger than that for $g=0$, and the deviation from the Fermi-liquid-like
behavior is realized.
In the case of $U=1.8$, where spin fluctuation is
strong, non-Fermi-liquid-like behavior is obtained as shown in Fig. \ref{fig5}(b).
For $g=0$, the value of $|S|$ is small and almost independent of $T$. On
the other hand, $|S|$ is drastically enhanced, and shows the peak at
$T^*\sim 150$K for $g=0.3$, where both the spin and orbital
fluctuation are strong.
The obtained result for $U=1.8$ and $g=0.3$ is consistent with experiments.\cite{Mcguire,Tropeano,Singh,Awana,Mun}
%Although the obtained $S$ is negative, which is consistent with
%experiments\cite{Mcguire,Tropeano,Singh,Awana,Mun}, the absolute value
%of $S$ for $g=0$ is much smaller than the result of experiments.
In the following, we explain why the absolute value of $S$ becomes large
for large value of $g$.
Since $v_{\a,\k}\sim 1/N_\a(\omega)$ at $\omega=\varepsilon^\a_\k$, where $N_\a$
and $\varepsilon^\a_\k$ are the
density of state and the dispersion on band $\a$, respectively, $S$ is rewritten as
\begin{eqnarray}
S&\propto&\frac{e}{\sigma_{xx}T} \sum_{\alpha}\int_{-\infty}^{\infty}
 d\omega \frac{z_\a\omega}{N_\a(\omega)\gamma_a(\omega)}\left(-\frac{\partial
						   f(\omega)}{\partial
						   \omega}\right)\nonumber\\
&\propto&\frac{-eT}{\sigma_{xx}}\sum_{\alpha} z_\a \frac{\partial}{\partial \omega}\left( \frac{1}{N_\a(\omega)\gamma_\a(\omega)}\right)_{\omega=0},
\end{eqnarray}
where $z_\a=\left(1-\frac{\partial \Sigma_\a(\omega)}{\partial
\omega}\right)^{-1}_{\omega=0}$ and $\gamma_\a$ are the renormalization
factor and the quasiparticle
damping on band $\a$, respectively.
When the orbital fluctuations are weak, $S$ takes small and negative
value because $\frac{\partial}{\partial \omega}N_\a>0$ is satisfied on
the e-pocket (cold spots). In the case of
strong orbital fluctuation with large value of $g$, $S$ is still negative while
the absolute value is much enhanced.
due to the large value of $\frac{\partial}{\partial \omega}\gamma_\a>0$ at
the cold spots on e-pocket shown in Fig. \ref{fig1}(a).
Thus, the orbital fluctuation plays an important role in enhancing the
absolute value of $S$ and reproducing the experimental results.
We stress that result for $U=1.8$ and $g=0.3$ well reproduce the
experimental behaviors of $S$ in optimum doped
Ba(Fe$_{1-x}$Co$_x$)$_2$As$_2$.\cite{Mun}

In the present FLEX approximation,
almost isotropic damping is obtained for $U=1.8$ and $g=0$, as shown in
Fig. \ref{fig1}(a).
In contrast, Kemper {\it et al}.\cite{Kemper} reported a clear hot/cold
spot structure due to the spin fluctuation
using the RPA self-energy (self-inconsistent FLEX approximation).
This difference would come from the presence (absence) of self-consistency
in the former (latter) calculation.
Kemper {\it et al}.\cite{Kemper} also reported interesting doping
dependence of the sign of Hall coefficient
$R_{\rm H}$. However, current vertex corrections would be necessary
to reproduce the magnitude and $T$-dependence of $R_{\rm H}$
appropriately.\cite{Kontani-RH}
%%%%%%%%%%%%%%%%%%%%%%%%%%%%%%%%%
\begin{figure}[!htb]
\includegraphics[width=0.9\linewidth]{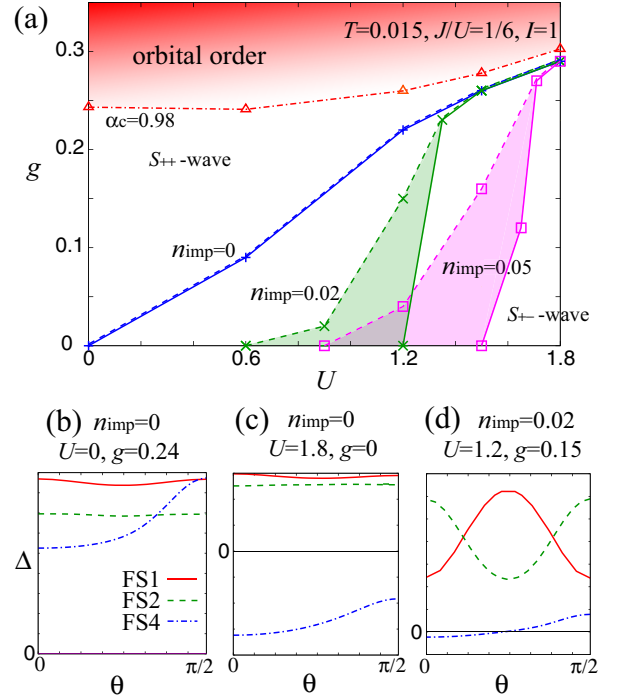}
\caption{(Color online) (a) $U$-$g$ phase diagram 
%for $T=0.015$ and $I=1$
 given by solving the linearized Eliashberg equation at $T=0.015$.
 Nodal $s$-wave gap state is obtained in the shaded area
 for $n_{\rm imp}=0.02$ and 0.05, and
 solid lines (dotted lines) represent the boundary between 
 fully-gapped $s_{+-}$-wave ($s_{++}$-wave) state.
 Dashed-dotted line denotes $\alpha_c=0.98$.
 (b) $s_{++}$-wave gap ($\lambda_E=0.59$)
 for $U=0$ and $g=0.24$.
 (c) $s_{+-}$-wave gap ($\lambda_E=0.49$)
 for $U=1.8$ and $g=0$ and.
 (d) Nodal $s$-wave gap ($\lambda_E=0.28$)
 for $U=1.2$ and $g=0.15$.
}
\label{fig2}

\end{figure}
%%%%%%%%%%%%%%%%%%%%%%%%%%%%%%%%%
\subsection{SC state}
Next, we discuss the SC state.
%based on the FLEX approximation.
In the presence of dilute impurities ($n_{\rm imp}\ll1$),
the linearized Eliashberg equation in the orbital basis is
\cite{Kontani-Onari}
\begin{eqnarray}
\lambda_{\rm E}\Delta_{ll'}(k)&=-&\frac{T}{N}\sum_{k',m_i}W_{lm_1,m_4l'}(k-k')
G'_{m_1m_2}(k')\nonumber\\
&\times&\Delta_{m_2m_3}(k')
G'_{m_4m_3}(-k')+\delta\Sigma^a_{ll'}(\epsilon_n),
\label{eqn:Eliash}
\end{eqnarray}
where $\Delta_{ll'}(k)$ is the gap function and
$\lambda_{\rm E}$ is the eigenvalue that reaches unity at $T=T_c$.
$\delta{\hat \Sigma}^a$ represents the impurity-induced gap function.
$(\hat{G'})^{-1}=(\hat{G})^{-1}-\delta\hat{\Sigma}^n$,
where $G$ is the Green function given by Eq. (\ref{eqn:Sigma}),
and $\delta{\hat \Sigma}^n$ is the impurity-induced normal self-energy.
The pairing interaction ${\hat W}$ in Eq. (\ref{eqn:Eliash}) is
\begin{equation}
\hat{W}(q)=\frac{3}{2}\hat{\Gamma}^s\hat{\chi}^s(q)\hat{\Gamma}^s-\frac{1}{2}\hat{\Gamma}^c\hat{\chi}^c(q)\hat{\Gamma}^c+\frac{1}{2}\hat{\Gamma}^s-\frac{1}{2}\hat{\Gamma}^c,
\label{eqn:W}
\end{equation}
where $\hat{\chi}^{s,c}$ is given by the FLEX approximation for $n_{\rm imp}=0$, 
because of the fact that the fully self-consistent FLEX with impurity-induced 
self-energy leads to unphysical reduction in $\chi^s$,
unless vertex correction is taken into account \cite{ROP}.
The first (second) term in Eq. (\ref{eqn:W}) works to 
set $\Delta_{\rm FS1,2}\cdot \Delta_{\rm FS3,4}<0$ ($>0$).

In the $T$-matrix approximation,
$\delta{\hat \Sigma}^{n,a}$ is given as
\begin{eqnarray}
\delta\Sigma^n_{ij}(\epsilon_n)&=&n_{\rm imp}T_{ij}(\epsilon_n) ,
 \label{eqn:dSn} \\
\delta\Sigma^a_{ij}(\epsilon_n)&=&n_{\rm imp}\sum_{lm}T_{il}(\epsilon_n)
f_{lm}(\epsilon_n)T_{jm}(-\epsilon_n)
 \label{eqn:dSa},
\end{eqnarray}
where 
$T_{ij}(\e_n)\equiv I(1- I{\hat g}(\e_n))^{-1}$
is the $T$-matrix in the normal state \cite{Onari-impurity};
${\hat g}(\e_n)\equiv \frac1N\sum_\k{\hat G}_\k(\e_n)$
is the local normal Green function, and $I$ is the local 
impurity potential that is diagonal in the orbital basis.
We put $I=1$ hereafter.
In Eq. (\ref{eqn:dSa}), 
$f_{ij}(\epsilon_n)=\frac1N\sum_{\bm{k},lm}G_{il}(k)\Delta_{lm}(k)G_{jm}(-k)$
is the linearized local anomalous Green function.

In Fig. \ref{fig2}(a), we show the $U$-$g$ phase diagram 
obtained by the FLEX approximation.
The dashed-dotted line represents the condition $\alpha_c=0.98$
at $T=0.015$, corresponding to $g=0.25\sim0.3$.
(In the RPA, the same condition 
is satisfied for $g=0.21\sim0.23$. \cite{Saito})
Therefore, substantial orbital fluctuations emerge
for $\lambda=gN(0)\lesssim0.2$
even if the self-energy correction is taken into account.
On the other hand,
$\alpha_s=0.95$ (0.92) for $U=1.8$ and $g=0$ (0.3)
in the FLEX approximation,
although $U_{\rm cr}=1.25$ for $\alpha_s=1$ in the RPA.
Thus, the renormalization in $\a_s$ is much larger than that in $\a_c$,
because of the difference in the coefficients (in factor 3)
between the first and the second terms in Eq. (\ref{eff}).

Next, we solve Eq. (\ref{eqn:Eliash}) with high accuracy
using the Lanczos method at $T=0.015$.
Then, the $s_{++}$-wave gap function is obtained around the line $\alpha_c=0.98$;
Figure \ref{fig2}(b) shows the $s_{++}$-wave gap 
for $g=0.24$ and $U=0$ ($\lambda_E=0.59$).
On the other hand, $s_{\pm}$-wave gap is obtained when 
$g$ is sufficiently small;
Figure \ref{fig2}(c) shows the $s_{\pm}$-wave gap 
for $U=1.8$ and $g=0$ ($\lambda_E=0.49$).  
When $n_{\rm imp}=0$, the gap function changes 
from (b) to (c) discontinuously on the phase boundary in Fig. \ref{fig2}(a),
as found in Ref. \cite{Saito}.
When $n_{\rm imp}\ge0.02$, however, the
gap function changes smoothly during the crossover.
Then, line-nodes inevitably appear on FS3,4 in the shaded area
in Fig. \ref{fig2}(a);
Figure \ref{fig2}(d) shows the nodal $s$-wave gap 
for $U=1.2$, $g=0.15$ and $n_{\rm imp}=0.02$ ($\lambda_E=0.28$).
Thus, both regions for $s_{++}$-wave and nodal $s$-wave states are extended 
%when $n_{\rm imp}\ne0$,
by a small amount of impurities, 
although $\lambda_E$ for the latter state is reduced by impurities.
A nodal $s$-wave solution at $n_{\rm imp}=0$ with larger $\lambda_E$
may be obtained by considering a 3D 
nodal-line structure in a 3D tight-binding model \cite{Mazin2}.

Here, we discuss that line nodes originate from
the competition between the orbital and spin fluctuations:
The electrons at $\theta\sim0$ ($\pi/2$) on FS4
are composed of orbital 2,3 (4).
Since the orbital 4 is absent in FS1,2,
the nesting-driven spin correlation between 
the orbital 2,3 on FS1,2 and the orbital 4 on FS3,4 is weak.
(That is, $\chi^s_{24,42}(\Q)\ll\chi^s_{22,22}(\Q)$.)
On the other hand, both $\chi^c_{24,42}(\q)$ and $\chi^c_{22,22}(\q)$
develop well \cite{Kontani-Onari,Saito,Saito2}.
Therefore, when orbital and spin fluctuations are comparable,
$\Delta_{\rm FS1,2}\cdot \Delta_{\rm FS4}$ is negative (positive) 
at $\theta\sim0$ ($\pi/2$)
due to the orbital-dependences of the spin and orbital susceptibilities.

%%%%%%%%%%%%%%%%%%%%%%%%%%%%%%%%%
\begin{figure}[!htb]
\includegraphics[width=0.99\linewidth]{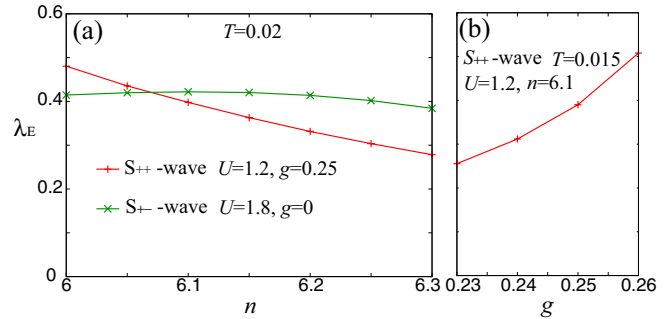}
\caption{(Color online) 
(a) $n$ dependence of $\lambda_{\rm E}$ for
 $s_{++}$- and $s_{+-}$-wave states at $T=0.02$ and $n_{\rm imp}=0$. 
(b) $g$ dependence of $\lambda_{\rm E}$ for $s_{++}$-wave state
at $T=0.015$ and $n_{\rm imp}=0$. 
}
\label{fig3}
\end{figure}
%%%%%%%%%%%%%%%%%%%%%%%%%%%%%%%%%

In Fig. \ref{fig3}(a), we show the filling dependence of $\lambda_{\rm E}$
for the $s_{++}$-wave state ($U=1.2$, $g=0.25$), and that 
for the $s_{+-}$-wave state ($U=1.8$, $g=0$).
We note that FS1,2 disappear for $n>6.3$. 
The value of $\lambda_{\rm E}$ for the $s_{++}$-wave state decreases 
monotonically with $n$, while $\lambda_{\rm E}$ for the $s_{+-}$-wave state 
is rather insensitive to $n$, 
maybe because the temperature, $T=0.02$, is rather high.
Figure \ref{fig3}(b) shows that $\lambda_{\rm E}$ 
for the $s_{++}$-wave state ($U=1.2$, $n=6.1$)
increases with $g$. 
%due to the increase of the orbital fluctuation.

\section{Conclusion}
We performed the FLEX approximation in the 
multiorbital Hubbard model including the charge quadrupole interaction for iron pnictides.
It was confirmed that the orbital-fluctuation-mediated $s_{++}$-wave state
is realized by small $e$-ph interaction $g$.
%As the orbital fluctuations due to $e$-ph interaction develop,
As increasing the value of $g$, both the $T_{\rm c}$ of the $s_{++}$-wave state and 
the resistivity $\rho$ are increased, 
and the latter changes from $T$-concave to $T$-convex.
This correlation between $T_{\rm c}$ and $\rho$ is
consistent with experiment \cite{Eisaki}.
Moreover, the obtained thermo-electric power $S$ is a large negative
value due to cold
spots on the e-pocket, when the
orbital fluctuation is dominant.
The large negative value of $S$ is consistent with experiments\cite{Mcguire,Tropeano,Singh,Awana,Mun}.
%This non-Fermi-liquid-like behavior typical in high-$T_{\rm c}$ samples
%is consistent with experiments \cite{Eisaki}.
We note that the region of $s_{++}$-wave or nodal $s$-wave states is enlarged
in the presence of a small amount of impurities.
Thus, the present orbital fluctuation scenario presents a unified 
explanation for both normal and SC electronic states.

\acknowledgements
We are grateful to M. Sato, Y. Kobayashi, Y. Matsuda, T. Shibauchi,
D.S. Hirashima, Y. Tanaka, K. Yamada, and F.C. Zhang 
for valuable discussions. 
This study has been supported by Grants-in-Aid for Scientific 
Research from MEXT of Japan, and by JST, TRIP.

%%%%%%%%%%%%%%%%%%%%%%%%
%references
%%%%%%%%%%%%%%%%%%%%%%%%

\end{document}